\documentclass[10pt]{article} 
\usepackage{amsmath,amssymb,cite,color,enumerate,epsfig,graphicx,fullpage,hyperref}
\usepackage{amsthm,changes}
\hypersetup{colorlinks=true,linkcolor=black,citecolor=black,urlcolor=blue}

\begin{document}
\title{A mathematical framework for predicting lifestyles of viral pathogens}
\author{Alexander Lange\,$^{1,2}$\\[1ex]
$^1$\,Computational Systems Biology, Zuse Institute Berlin, Germany\\
$^2$\,Department~of Applied~Biosciences~and Process Engineering, Anhalt UAS\\[1ex]
E-mail: lange@zib.de}
\date{}

\maketitle

\begin{abstract}

Despite being similar in structure, functioning, and size viral pathogens enjoy very different mostly well-defined ways of life. They occupy their hosts for a few days (influenza), for a few weeks (measles), or even lifelong (HCV), which manifests in acute or chronic infections. The various transmission routes (airborne, via direct contact, etc.), degrees of infectiousness (referring to the load required for transmission), antigenic variation/immune escape and virulence define further pathogenic lifestyles. To survive pathogens must infect new hosts; the success determines their fitness. Infection happens with a certain likelihood during contact of hosts, where contact can also be mediated by vectors. Besides structural aspects of the host-contact network, three parameters/concepts appear to be key: the contact rate and the infectiousness during contact, which encode the mode of transmission, and third the immunity of susceptible hosts. From here, what can be concluded about the evolutionary strategies of viral pathogens? This is the biological question addressed in this paper. The answer extends earlier results (Lange \& Ferguson\- 2009, PLoS Comput Biol 5 (10): e1000536) and makes explicit connection to another basic work on the evolution of pathogens (Grenfell et al.~2004, Science 303: 327--332). A mathematical framework is presented that models intra- and inter-host dynamics in a minimalistic but unified fashion covering a broad spectrum of viral pathogens, including those that cause flu-like infections, childhood diseases, and sexually transmitted infections. These pathogens turn out as local maxima of the fitness landscape. The models involve differential- and integral equations, agent-based simulation, networks, and probability.

\end{abstract}

\section{Introduction}

In the light of the many incurable and newly emerging viral infections, such as HIV, HCV, pandemic influenza, dengue, SARS or Ebola, to mention a few, one is interested in knowing more about possible ways viruses can exist in the human host population.
By employing numerical models we are trying to learn about their evolutionary strategies and how these strategies depend on the host environment.

Due to the complexity of viral habitats---often located within several host species---and due to the various transmission routes between hosts,
\footnote{which can involve special environmental conditions (e.g., temperature \cite{HBSR13})}
there is no consistent mathematical
framework for studying more general virus-related questions.
Most of the literature studies particular infections
\cite{MMP13,FLLSHAB14}
and often either focuses on between- \cite{FHCWH07} or on within-host dynamics \cite{ALR11,JKAA12,HLSR14}.
Some papers do follow a more general approach, e.g.,
combine inter- and intra-host dynamics \cite{CGB07,LA09,PVBWG10},
discuss involved challenges \cite{HR15,GPM_15,LFMRW15},
or sketch a unified perspective \cite{GPGWDMH04,LF09}.
The two last mentioned ones are of particular interest here.

Do they cover the same predictions? This is the question we would like to answer.
Translation between different frameworks is usually not straightforward.
Therefore, our first goal aims at establishing interpretation: we want to re-identify concepts from \cite{GPGWDMH04} within the framework of \cite{LF09}.
In particular, we try to relate the so-called {static patterns} of \cite{GPGWDMH04} and the {infection types} of \cite{LF09}.

Besides the mathematical methodology, a crucial part of any modeling framework is the involved parameters, referring to their number, meaning, and importance.
Therefore, we try to compare the parameters of the two chosen approaches.
We aim to reconstruct the infection types of \cite{LF09} by suggesting a minimal set of parameters, and we aim to mathematically formulate the evolutionary strategies
behind.

Similar to other forms of life, the evolutionary success of viruses correlates with their success to replicate. To take account of that, we study viral replication within- and between hosts.
Similar to the methods used in \cite{LF09}, we employ differential- and integral equations, stochastic models, and numerical simulations.
Based on the various parameter sets that are involved, we investigate conditions that maximize the reproductive fitness.
But before going deeper into that, we briefly recall aspects of the two frameworks, \cite{GPGWDMH04} and \cite{LF09}, that are important here.

\section{Background}

\subsection{The phylodynamic framework \label{sect_G04}}

\begin{figure}\hspace{15em}
\includegraphics[width=17em]{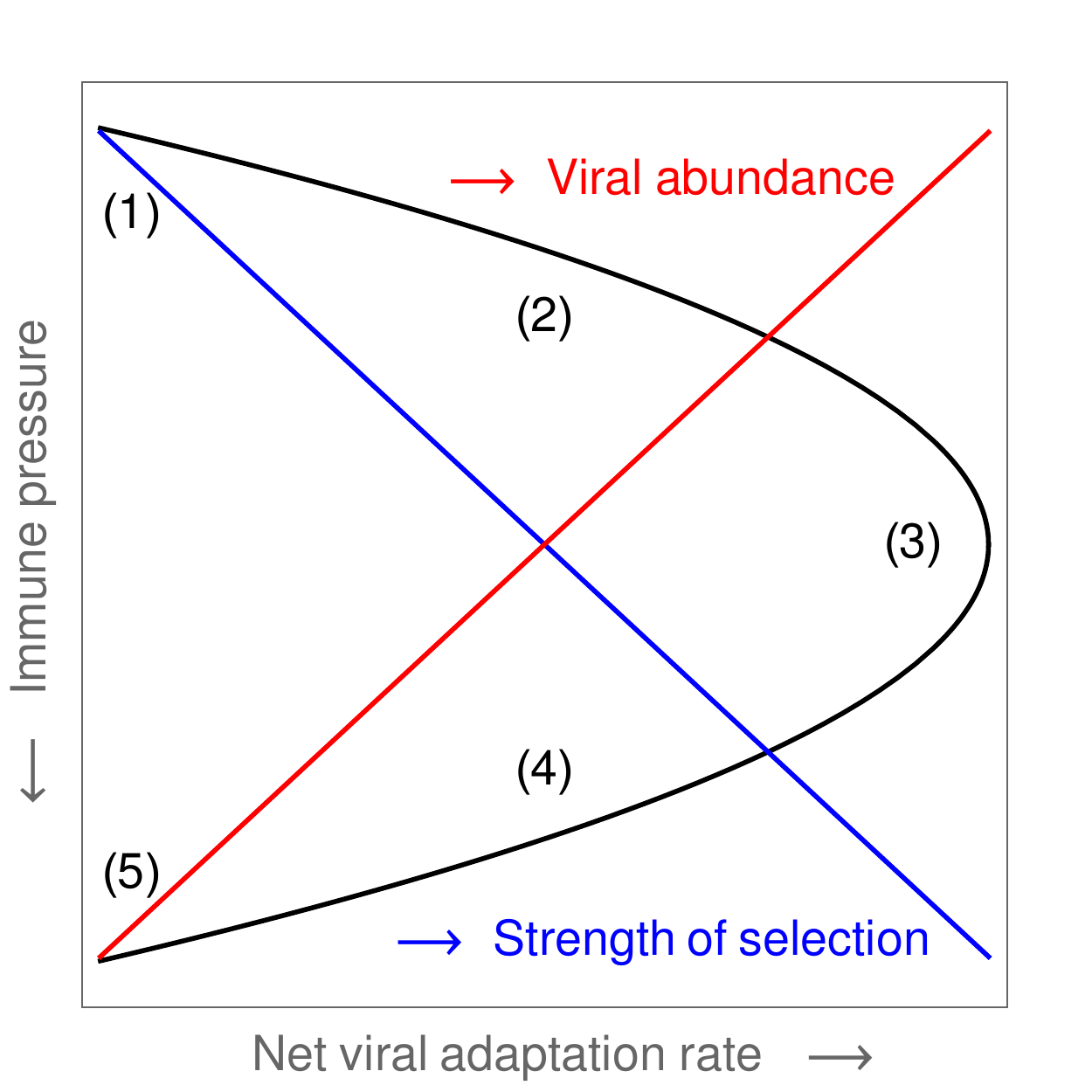}
\caption{\label{fig_G04}
Static patterns. The figure is a $90^\circ$-rotated sketch of Figure~2A in \cite{GPGWDMH04}. It indicates the locations of the five {\em static patterns} (lying on a ``parabola'') in the pathogen parameter space of Grenfell et al.~\cite{GPGWDMH04}, which is formed by the {\em immune pressure} and the {\em net viral adaptation rate}. Furthermore, the figure indicates the monotonic behavior of the {\em strength of selection} (blue) and the {\em viral abundance} (red) with respect to the immune pressure (y-axis).
}
\end{figure}

Well-known for coining the term {\em phylodynamics}, the paper by Grenfell et al.~\cite{GPGWDMH04}
suggests five so-called {\em static patterns} to characterize ``types'' of pathogens (Fig.~\ref{fig_G04}).
The following identifications and examples (using RNA-viruses) are given:

\begin{itemize}
\item[(1)] no effective immune response, no adaptation
(HCV in immuno-compromised hosts,\\ influenza A virus immediately after an antigenic shift);
\item[(2)] low immune pressure, low adaptation (rapidly progressing chronic HCV and HIV);
\item[(3)] medium immune pressure, high adaptation (antigenic drift in influenza A virus,\\ intra-host HIV infections);
\item[(4)] high immune pressure, low adaptation (HIV in long-term non-progressive hosts);
\item[(5)] overwhelming immune pressure, no adaptation (measles and other morbilliviruses).
\end{itemize}

\subsection{Transmission mechanisms and viral evolution}

\begin{figure}
\hspace{5em}
\includegraphics[width=35em]{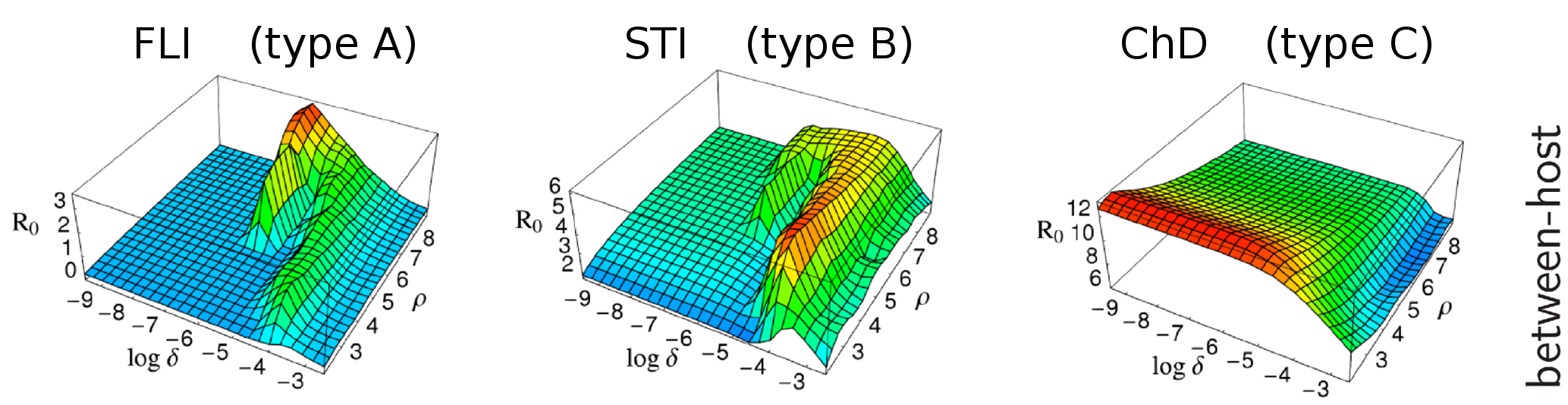}\\[2ex]
\hspace*{5em}
\includegraphics[width=35em]{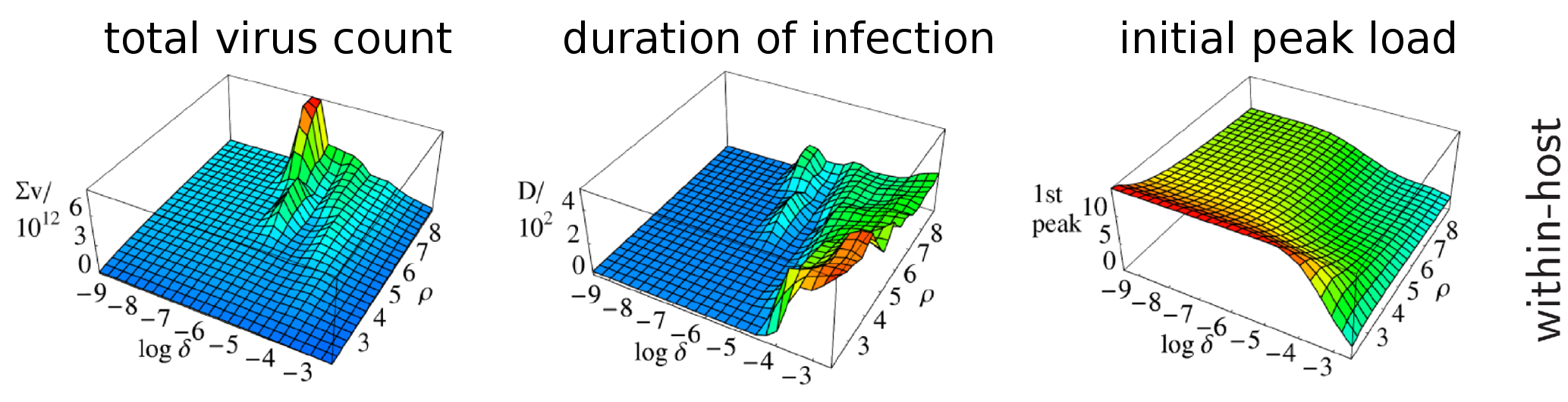}
\caption{\label{fig_LF}
Infection types. This figure is adopted from Figure 3 in \cite{LF09}.
The top row shows the fitness landscapes (due to between-host replication, $R_0$) over pathogen space ($=$ {\em antigenic variation} $\times$ {\em intra-host replication}) for flu-like infections (FLI), sexually transmitted infections (STI), and childhood diseases (ChD).
The bottom row shows the corresponding between-host characteristics: total virus count, duration of infection, and the initial peak load, respectively.
The maxima of these surfaces define three evolutionary strategies (or {\em lifestyles}, as we also refer to them). 
While having the maxima at the same location in pathogen space, the surfaces of the top and bottom rows are similar too.
}
\end{figure}

The work by Grenfell et al. focuses on the viral population and the host-immune response.
Their approach is rather independent from epidemiological aspects
such as transmission and inter-host environment.
This is different in \cite{LF09},
where infectious diseases are classified into three types
(cf.~Fig.~\ref{fig_LF}).
Even if the classification is based on antigenic variation (being either A: medium, B: high, or C: low), epidemiological aspects such as the host-contact rate and the transmission mode are revealed to be closely related.
Each infection type corresponds to a certain range of contact/transmission rates (A: low, B: medium, C: high).
Depending on that range, each infection type shows a distinct fitness landscape (between-host reproduction) over pathogen space (Fig.~\ref{fig_LF}, top row).
Most interestingly, the infection types correspond to three evolutionary strategies
(Fig.~\ref{fig_LF}, bottom row):
\begin{align}
\begin{cases}
A\\B\\C 
\end{cases}
\text{maximizes the}\quad
\begin{cases}
\text{total viral load ,}\\
\text{duration of infection ,}\\
\text{initial peak load ,}
\end{cases}
\end{align} 
where, to some extend, the fitness landscapes resemble the strategic ones (top and bottom raws in Fig.~\ref{fig_LF}).

\section{Methods \label{sec_meth}}

\begin{figure}\hspace{8em}
\includegraphics[width=30em]{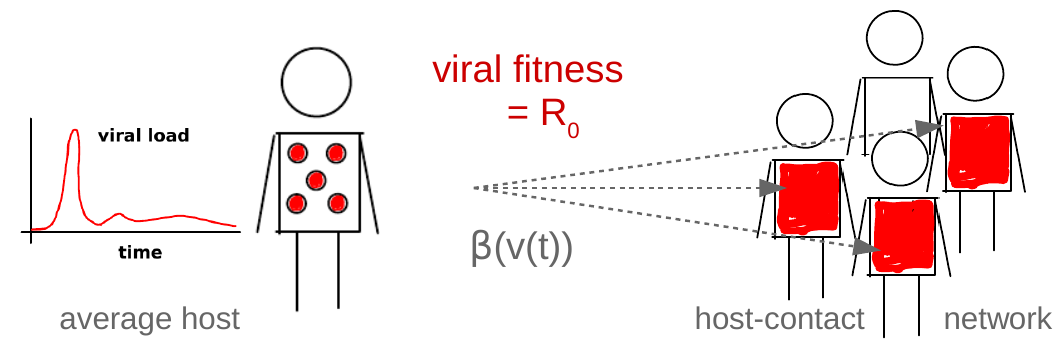}
\caption{\label{fig_fitness}
Modeling framework.
Systematically, for all viruses from our pathogen parameter space,
we simulate the within-host evolution
and calculate the average load over time $v(t)$.
The load curve is used to define a time-dependent transmission rate, $\beta(v(t))$.
Based on this rate, the between-host dynamics is simulated for a totally susceptible host-contact network. The total number of infected individuals then determines the basic reproduction number $R_0$ and hence viral fitness.
}
\end{figure}

We study a highly simplified scenario of viral replication
that includes intra- and inter-host dynamics (cf.~Fig.~\ref{fig_fitness}).
The link between the two is established by a transmission model, which leads to quantifying viral fitness
in terms of the basic reproduction number.
The intra-host model involves cells for viral replication and an adaptive immune response.
Via mutations, viral replication includes a stochastic element.
The simulation outcome represents the viral load of an average host.
While, for simplicity, all host individuals are considered equal, our inter-host model does involve structure of a contact/transmission network.

\subsection{Viral fitness\label{sec_vf}}

In an inter-host context, viral fitness is measured by the success of the virus to reproduce while reaching new hosts.
This includes viral reproduction within hosts and transmission to other hosts.
The latter, formalized by the basic reproduction number,
defines the mathematical concept that we utilize for predicting viral fitness \cite{AM82}.
Implicitly---via the viral load (cf.~Sect.~\ref{sect_td} below)---this number also takes account of the reproduction within hosts.

The basic reproduction number $R_0$ counts the infections in a totally susceptible population
that are caused directly by one infective individual. 
That is, to determine $R_0$ one must study the contact neighborhood of an infected individual,
which initially only contains susceptibles, $S(0)=N-1$. 

When modeled by the mass-action law, 
the growth of the number of infections resulting from one infective individual, $I(0)=1$, is given by
$I'(t)=\beta(t)\,S(t)\,I(0)$. Integration then yields $R_0$.
In practice, one must introduce a cut-off as an upper time limit, 
\begin{align}\label{eq_R0}
R_0=\int_0^D\beta(t)\,S(t)\,dt\,.
\end{align}
In our simulations, this cut-off is modeled by the first entering time,
$D=\inf\{t>0\,|\,v(t)\le v_0\}$, 
\footnote{$D$ usually turns out to be shorter than 2 years.}
capturing the time (referred to as {\em duration of infection}) when the viral load $v(t)$ falls below a 
critical value $v_0$.

Important to note that we do not explicitly consider intermediate hosts or vectors here, but neither we exclude them. Mass-action can provide an effective description for vectors \cite{L16}.

\begin{table}
\center
\begin{tabular}{l|l}
Symbol & Parameter \\\hline
$c_0$ ($10^8$) &initial/max resource\\
$v_0$ ($10$) & initial/min viral load\\
$x_0$ ($1$)& initial/min immunity\\
$\widehat\alpha$ ($10^{-5}$) & upper infectiousness bound\\
$\gamma$ ($1$) & replenishment of resource\\
$\varepsilon$ ($0.25/\sigma$)& innate immunity\\ 
$\zeta$ ($0.8$) & growth of immunity\\
$\eta$ ($10^3$) & saturation of immunity\\
$\mu$ ($0.1$) & mutation rate\\
$\nu_1$ ($10^3$) & virions per resource unit\\
$\xi$ ($0.3$)& decline of immunity\\
$\sigma$ ($10^{-3}$) & clearance due to immunity \\
$\varphi$ ($0.25$) & cliquishness\\
$\chi$ ($0.4$)& cross-immunity\\
\end{tabular}
\caption{\label{t_fixed_para} Fixed parameters. Values we used are given in brackets. Time units are always days.}
\end{table}

\subsection{Intra-host model}

\begin{figure}
\hspace*{1.75em}(a)
\includegraphics[height=52mm]{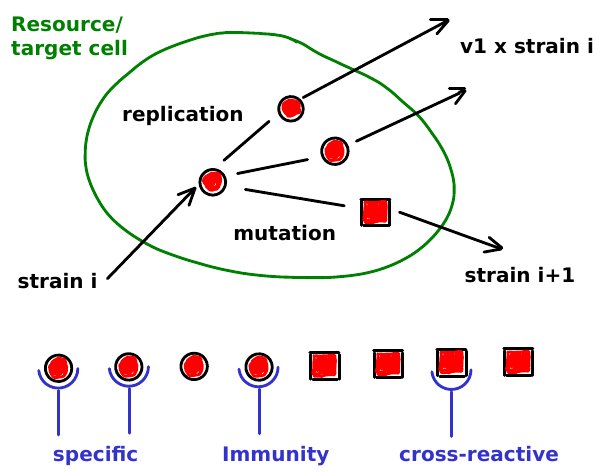}
\hspace*{.75em}(b)
\hspace{.75em}
\includegraphics[height=26mm]{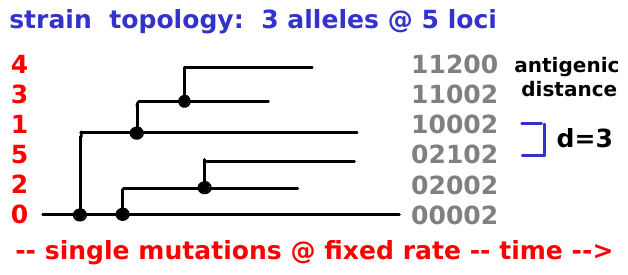}\\
\caption{\label{fig_mutations_loci-alleles}
Within-host replication. Panel (a) illustrates the replication of viral strain $i$ into multiple identical copies (about $\nu_1$) and the mutant strain $i+1$. Before specific immunity develops, there is a cross-reactive immune response from the earlier strain $i$. Cross-reactive immunity is exerted based on the loci-allele structure indicated in Panel (b).
Its strength depends on the antigenic distance between the involved strains; cf.~Eq.~\eqref{y_i}. The distance is associated with the number of mutations required to transform one strain into the other.}
\end{figure}

For the viral dynamics within the host, we apply one of the simplest compartmental models \cite{LF09} that involves multiple viral strains, adaptive immune responses, and target cells that provide the resource for viral replication; see Figure~\ref{fig_mutations_loci-alleles}a.
In part, replication is assumed to lead to mutations
(governed by a Poisson process of rate $\mu\rho$)
and to the creation of novel strains (at frequency $\delta$).
\footnote{The vast majority of mutations is assumed to be detrimental to the virus.}
The antigenic appearance of the virus (modeled through a loci-allele structure
as illustrated in Fig.~\ref{fig_mutations_loci-alleles}b)
varies between different strains.
\footnote{Mutations are not supposed to change intra-host parameters except for $\delta,\rho$. This is a reasonable assumption for short time scales.}
Primarily, immunity is directed towards one specific strain,
although it is assumed to provide cross-protection from other antigenically close strains.
Mathematically, the immune response (towards strain $i$) is modeled via a function,
\footnote{$[a]_+=(|a|+a)/2$ denotes the positive part of the term $a$}
\begin{align}\label{y_i}
y_i(x)=\sum_{k\le n} x_k\cdot\left[1-(1-\chi)\,\varrho_{ik}\right]_++\varepsilon>0\,,
\end{align}
that accumulates all the available amounts $x_k$ of specific immunity
weighted by the antigenic distance 
($\varrho_{ik}=\#\text{ non-coinciding loci of strains $i$ and $k$}$;
cf.~Fig~\ref{fig_mutations_loci-alleles}b).
This function depends on a cross-immunity parameter $\chi\in[0,1]$;
in this paper it is supposed to cover innate immunity $\varepsilon$ as well.

Between mutation events that lead to novel strains,
\footnote{Novel strains $j$, produced by a Poisson process of rate $\delta\mu\rho$, are introduced by a set of two new equations of index $j$ and initial values (e.g., $v_j(0)=10$, $x_j(0)=1$).}
the time evolution of viral loads $v_i$, of specific immunity $x_i$, and of target cells $c$
is modeled by a system of asymptotically linear ODEs,
\begin{subequations}
\begin{align}
\frac{dv_i}{dt}&=(1-\mu)\,\rho\,v_i^+(c)
-\sigma\,v_i^-(x)\,,\\
\frac{dx_i}{dt}&=\xi\,(x_0-x_i)+\zeta\,x_i^+(v_i)\,,\\
\frac{dc}{dt}&=\gamma\,(c_0-c)-\rho\,c^-(v)\,.
\end{align}
\end{subequations}
The response to the virus is based on the following interaction terms (that model)
\begin{subequations}
 \begin{align}
v_i^+(c)&=v_i\cdot h_{v/\nu_1}(c)
&&\text{\hspace{-3em}(replication of strain $i$ depending on the available target cells)}\,,\\
v_i^-(x)&=v_i\cdot y_i(x)
&&\text{\hspace{-3em}(removal of strain $i$ due to the immune response)}\,\,,\\
x_i^+(v_i)&=x_i\cdot h_\eta(v_i)
&&\text{\hspace{-3em}(activation of specific immunity to strain $i$)}\,,\\
c^-(v)&=c\cdot h_{c}(v/\nu_1)
&&\text{\hspace{-3em}(target cell depletion due to infection)}\,;
\end{align}
\end{subequations}
the involved rates are listed in Table \ref{t_fixed_para}. 
Hill functions $h_a(b)=\frac{b}{a+b}\in[0,1]$ are employed
to scale the virus production according to the available target cells
and to implement a load-dependent immune response.
Target cell depletion is derived entirely from virus production,
$c^-=\frac{1}{\nu_1}\sum_iv_i^+$.
The resulting interaction terms behave linearly,
\begin{subequations}
 \begin{align}
v_i^+(c)&=v_i&&\text{\hspace{-9em}if\quad $c\gg v/\nu_1$\quad(target cell number is large)}\,,\\
x_i^+(v_i)&=x_i&&\text{\hspace{-9em}if\quad $v_i\gg\eta$\quad(strain-specific viral load is high)}\,,\\
c^-(v)&=c&&\text{\hspace{-9em}if\quad $v\gg\nu_1c$\quad(viral load is high)}\,.
\end{align}
\end{subequations}
Under opposite conditions, each of these terms vanishes.
In particular, $v_i^+(c)=0$ if $c\ll v/\nu_1$, which reflects saturation effects caused by the limited number of target cells.
In the virus-free equilibrium, all the interaction terms vanish and the system of ODE decouples: $v_i=0$, $x_i=x_0$, $c=c_0$.

\subsection{Transmission dynamics\label{sect_td}}

According to our fitness definition, we need to study viral transmission between hosts.
Here we assume that 
the rate of transmission
depends on the viral load $v$ of the transmitting (average) host.
A simple model is given by an exponential law,
\begin{align}
\beta=\widehat\beta\cdot(1-e^{-\alpha\,v})
\end{align}
where $\alpha$ represents a load-dependent infectiousness parameter and $\widehat\beta$ the load-saturated transmission rate.
This coefficient,
\begin{align}
\widehat\beta=\frac{\kappa\,\lambda}{N}\,,
\end{align}
which is taken with respect to a reference population,
is formed by the contact rate $\kappa$, the likelihood $\lambda$ of transmission per contact, and the average number $N$ of individuals in the contact neighborhood of a single host.
Typical parameter values are given in Table~\ref{T_contact_rate}.
Together, the parameters $\alpha$ and $\widehat\beta$ encode the mode of transmission.

As a consequence of the within-host dynamics and the time-dependent viral load $v(t)$, 
the transmission rate is also a function of time, $\beta(t)$. Its initial value corresponds to the transmitted viral load at the time of infection, $t=0$.

\begin{table}
 \begin{center}
\begin{tabular}{c|c||c|c|c||c|c||c}
Infection & type&$\kappa$&$\lambda$&$N$&$\widehat\beta$&$\mu$&$R_0$\\\hline\hline
ChD & C &9&0.4&10&0.36&0.024&15\\\hline
STI & B &0.5&0.6&4&0.08&0.016&5\\\hline
FLI & A &15&0.1&50&0.03&0.015&2
\end{tabular} 
\end{center}
\caption{Transmission parameters. Exemplary values for childhood diseases (ChD) / type C infections, sexually transmitted (STI) / type B infections, and flu-like (FLI) / type A infections.
\label{T_contact_rate}}
\end{table}

\subsection{Host network}

The viral dynamics between hosts is modeled most realistically on a network, where potential hosts represent the nodes linked to each other via potential contacts. A particular fraction of contacts ($\lambda$, specific to the infection) transmits the virus from one to another host.
To quantify the reproductive fitness of the virus, we study the transmission network only for the contact neighborhood of one initially infected host.
For this neighborhood, we determine the number of susceptibles over time and
eventually calculate the basic reproduction number.

\begin{figure}\hspace{16em}
\includegraphics[width=13em]{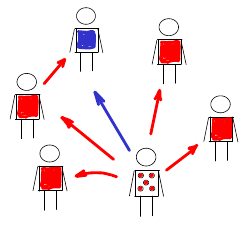}
\caption{\label{fig_screening}
Screening effect. The Sketch illustrates the reduced number ($4<5$) of susceptibles (red) in the contact neighborhood of one infected individual (red dots), caused by one secondary infection in the host network ($\varphi\neq0$). The {\em screened\,} individual (blue) cannot be infected by the initially infected individual anymore. This reduces the basic reproductive number in comparison to an idealistic network-free scenario ($\varphi=0$).
}
\end{figure}

Different from a simple mass-action model,
the mathematical formalism describing a network incorporates a cliquishness parameter $\varphi$,
which quantifies the number of contacts between members of the considered network-neighborhood.
These network contacts help spreading the virus and, as a consequence, effectively lower the number of susceptibles in the neighborhood.
This phenomenon is referred to as screening effect (cf.~Fig.~\ref{fig_screening}).

In the contact neighborhood of the initially infected host, the spread of the virus can be described in terms of two compartments, representing susceptible $S$ and infective individuals $I$.
The generation of infected individuals (at time $t$) is given by
\begin{subequations}\label{seq_I}
 \begin{align}\label{eq_I1}
I'(t)=\beta(t)\,S(t)&+\varphi~S(t)\int_0^td\tau_1\,\beta(t-\tau_1)\,I'(\tau_1)\\
&+\varphi^2\,S(t)\int_0^{t}d\tau_2\,\beta(t-\tau_2)\,I'(\tau_2)
\,S(\tau_2)\int_0^{\tau_2}d\tau_1\,\beta(\tau_2-\tau_1)\,I'(\tau_1)+\cdots\,,
\end{align}
\end{subequations}
where the listed terms model transmissions from the initial host, secondary hosts (infected by the initial host), tertiary hosts (infected by secondary hosts), etc.
All these terms represent mass-action coupling.
Transmissions from secondary hosts are weighted by the network parameter $\varphi$, tertiary hosts by its square $\varphi^2$, etc.
The involved convolution products,
\begin{align}
(\beta\ast I')(s)=\int_0^sd\tau\,\beta(s-\tau)\,I'(\tau)=\int_0^s\beta(s-\tau)\,dI(\tau)\,,
\end{align}
provide load-weighted transmission rates (at time $s$, originating from new infections before $s$).
According to the mass-action law, these terms are multiplied by the numbers of susceptibles $S(s)$ in Eq.~\eqref{seq_I}.

To obtain an equation that only involves susceptibles, we replace $I'$ by $-S'$ based on the assumption that the size of the contact neighborhood of the initially infected host does not change over time,
\footnote{There is no need to introduce further compartments. Recovered individuals, for example, are modeled by infectives with low viral load.
However, one may include a small replacement of individuals in the contact neighborhood.
Its influence on possible infection types has been discussed in \cite{LF09}.}
\begin{align}\label{eq_sub}
N'=S'+I'=0\,.
\end{align}
The substitution is applied to Eq.~\eqref{seq_I}
and, to save computation time, only secondary hosts \eqref{eq_I1} are considered.
The resulting equation,
\begin{align}\label{eq_S'}
S'=-\left(\beta-\varphi\,\beta\ast S'\right)S\,,
\end{align}
which models the time evolution of susceptibles in the contact neighborhood of the initially infected,
is solved numerically starting with $S(0)=N-1$.
The resulting function, $S(t)$, is then used to calculate the basic reproduction number \eqref{eq_R0}.

\subsection{Fitness maxima}

In a setting that includes transmission between hosts,
the basic reproduction number adequately encodes viral fitness (cf.~Sect.~\ref{sec_vf}).
It is calculated here for two sets of parameters (two each),
$R_0(\widehat\beta,\alpha;\delta,\rho)$,
referred to as {\em pathogen space} $(\delta,\rho)$ and
{\em transmission space} $(\widehat\beta,\alpha)$.
These spaces are supposed to capture different ``types'' of viral pathogens.

\begin{figure}\hspace{8em}
\includegraphics[width=30em]{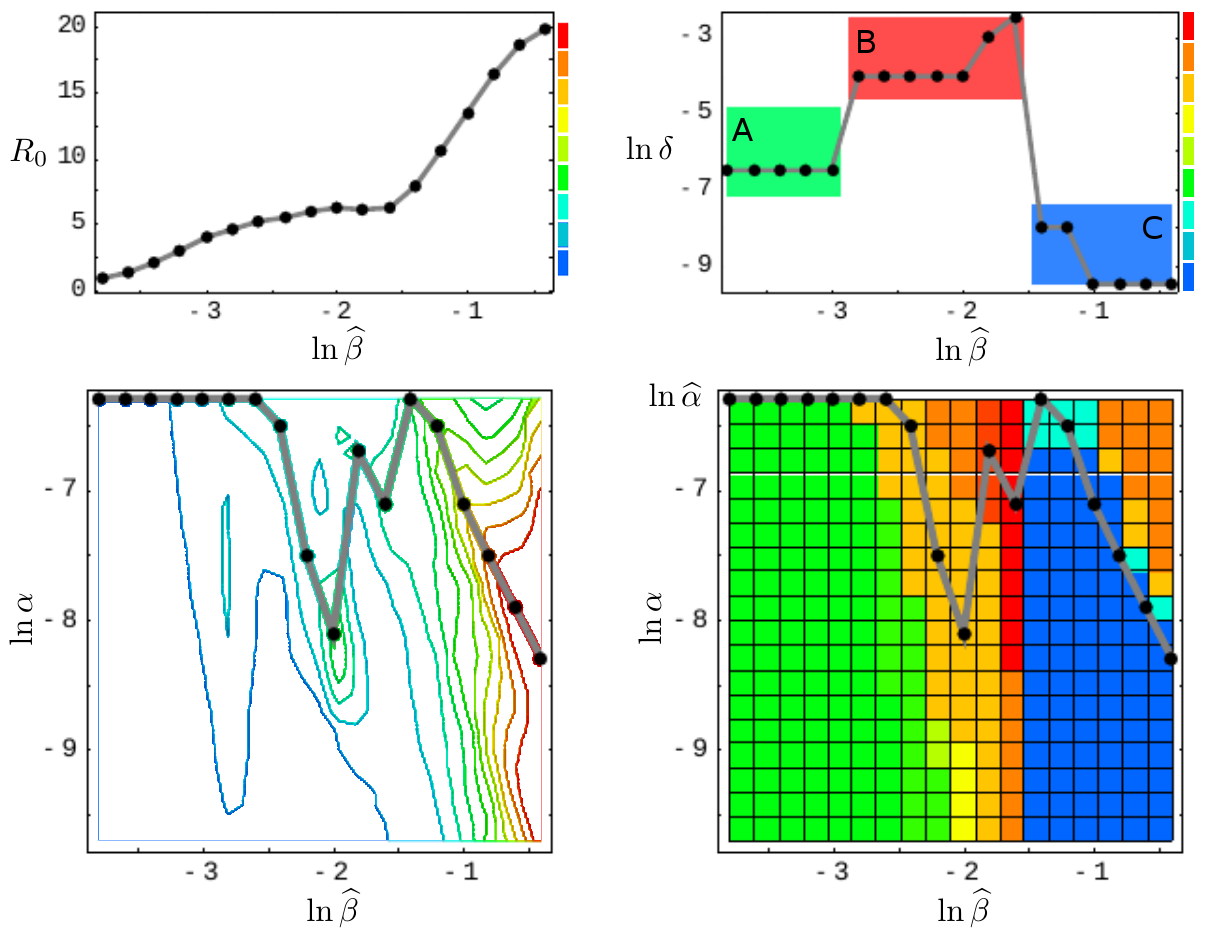}
\caption{\label{fig_max_fit}
Fitness and antigenic variation.
The figure illustrates the definition of the three infection types (A,\,B,\,C) based on antigenic variation (medium,\,high,\,low) and maximal fitness.
The lower left hand side panel shows the fitness landscape on transmission space, $\max_{\delta,\rho}\,R_0(\widehat\beta,\alpha;\delta,\rho)$. The corresponding top panel indicates the fitness maxima $\widehat R_0(\widehat\beta)$ for the simulated transmission rates $\widehat\beta$ (black dots).
The lower right hand side panel shows the antigenic variation ($\ln\,\delta$) over transmission space $(\widehat\beta,\alpha)$.
Here, the gray curve $\widehat\delta(\widehat\beta)$ selects the $\delta$-values that correspond to fitness maxima.
These $\delta$-values are shown in the corresponding top panel; they suggest a three-type classification.  
}
\end{figure}

In order to determine the types that are favored by evolution, one has to find parameter values,
\begin{align}
\widehat\delta(\widehat\beta)&=\underset{\delta}{\text{arg\,max}}\left(\max_{\alpha\le\widehat\alpha,\rho}\,R_0(\widehat\beta,\alpha;\delta,\rho)\right)\,,
\end{align}
as indicated for the antigenic variation (cf.~Fig.~\ref{fig_max_fit}), that maximize the viral fitness,
\begin{align}
\widehat R_0(\widehat\beta)&=\max_{\alpha\le\widehat\alpha}\,\max_{\delta,\rho}\,
R_0(\widehat\beta,\alpha;\delta,\rho)\,.
\end{align}
The antigenic variation is of particular importance. It offers a natural
classification leading to three infection types (referred to as A,B,C; cf.~Fig.~\ref{fig_max_fit}).

Variation of the four involved parameters is sufficiently general.
The cliquishness parameter $\varphi$, for example, as being another parameter, can be expressed
in \eqref{eq_S'}
by the neighborhood size, to which it is inversely related, $\varphi\propto 1/N$, approximately.
\footnote{This follows from the fact that simultaneous scaling of $S$ and $\varphi$ keeps Eq.~\eqref{eq_S'} invariant.}
Further parameters, such as cross-immunity $\chi$ and the infectiousness bound $\widehat\alpha$,
represent generic scenarios for a wide range of values.
They are kept fixed when deriving our first result. Their influence on the pathogenic lifestyle is investigated  afterwards, forming our second result.

\section{Results} 

Applying the model outlined above, one can straightforwardly reconstruct the static patterns of Grenfell et al.~\cite{GPGWDMH04}.
Furthermore, one can identify three parameters that---when adjusted appropriately---lead to the three infections types introduced in \cite{LF09}.
This is demonstrated in the following two subsections.

\subsection{Reconstruction of the static patterns \label{sec_r1}}

\begin{figure}\hspace{9em}
\includegraphics[width=33em]{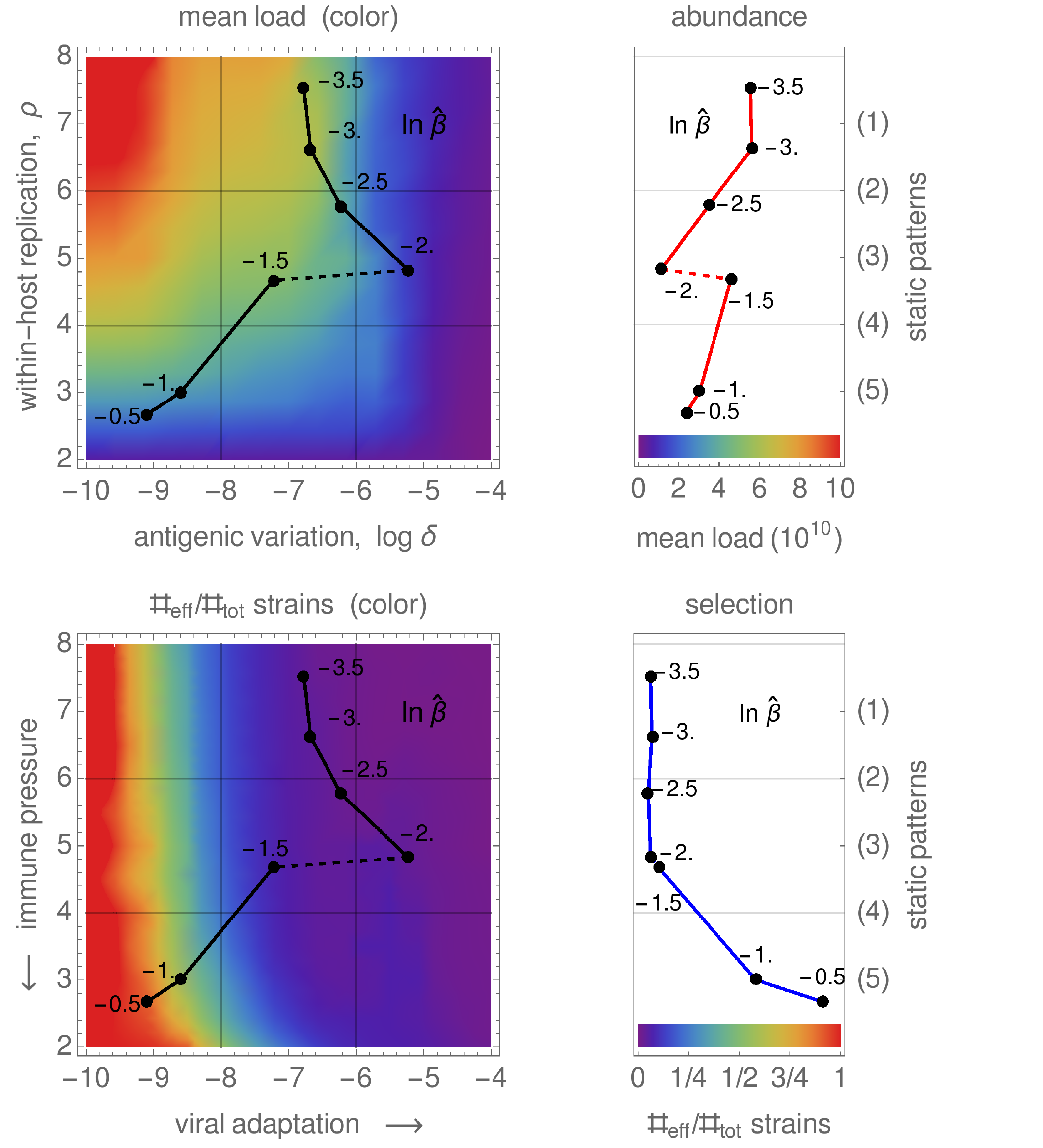}
\caption{\label{fig_static_patterns}
Reconstructed static patterns.
The top panels show the mean viral load (abundance) over the pathogen space (left) and the fitness maxima of several transmission rates (representing the five static patterns; right).
The bottom panels show the ratio of effective- to total strain numbers (representing the strength of selection) over the pathogen space (denoted as in \cite{GPGWDMH04}; left) and the fitness maxima of transmission rates (static patterns; right).
}
\end{figure}

We assume that the pathogen space of Grenfell et al.~\cite{GPGWDMH04} (cf.~Sect.\ref{sect_G04}) can be identified with ours via the following two correspondences,
\begin{subequations}
\begin{align}
\text{1~/~immune pressure}~&\sim~\text{intra-host reproduction, $\rho$}\,,\\
\text{net viral adaptation rate}~&\sim~\text{antigenic variation, $\delta$}\,,
\end{align}
\end{subequations}
where ``$\sim$'' encodes positive correlation.
Our first parameter, the intra-host reproduction, defines the reaction of the immune system to the virus, whereas our second parameter, the antigenic variation, already coincides with the one utilized by Grenfell et al.
By maximizing the basic reproduction number (Eq.~\eqref{eq_R0}) over these two parameters, and keeping all other parameters fixed,
\footnote{Here we refer to the infectiousness bound $\widehat\alpha$, cross-immunity $\chi$ and other intra-host parameters. In our numerical simulations, they are assigned with values from Table~\ref{t_fixed_para}.}
we obtain a $\widehat\beta$-depending curve that represents maximal values of viral fitness in pathogen space,
\begin{align}
\widehat\beta\mapsto
\underset{(\delta,\rho)}{\text{arg\,max}}~\widehat R_0(\widehat\beta)\,.
\end{align}
This curve (black, in left hand side panels of Fig.~\ref{fig_static_patterns}) resembles the parabola of Grenfell et al.~\cite{GPGWDMH04} (Fig.~\ref{fig_G04}), which defines five static patterns (cf.~the right hand side of Fig.~\ref{fig_static_patterns}).
We therefore hypothesize that the five patterns (numbered $1,\dots,5$) are positively correlated to the transmission rate $\widehat\beta$ (cf.~left hand side panels in Fig.~\ref{fig_static_patterns}).
In \cite{GPGWDMH04}, the five patterns have not been associated with inter-host concepts or a particular parameter.
Within our framework, the transmission rate $\widehat\beta$ happens to be a natural candidate in quantifying these patterns. By changing the value of $\widehat\beta$ one can shift
between patterns.
 
Furthermore, we are able to reconstruct the {\em viral abundance} and the {\em strength of selection} over the range of the static patterns (or, equivalently, the immune pressure; cf.~right hand side panels in Fig.~\ref{fig_static_patterns}). Here the following correspondences are employed,
\begin{subequations}
\begin{align}
\text{viral abundance}~&\sim~\text{mean viral load}~\, (=\bar v)\,,\\
\text{strength of selection}~&\sim~\text{ratio effective to total number of strains}~\,(=\#_\text{eff}/\#_\text{tot}) \,,
\end{align}
\end{subequations}
where the effective number of strains is associated with load-weighted strain-frequencies,
$\#_\text{eff}=\sum_i\bar v_i\#_i$, and $\#_\text{tot}=\sum_i\#_i$. 
For the viral abundance, we obtain a jump between the patterns 3 and 4 (or, equivalently, between $\ln\widehat\beta=-2$ and $-1.5$, as indicated by a dotted line in the top left panel of Fig.~\ref{fig_static_patterns}).
This discontinuity is visible as well in the maximized fitness curve on the left hand side panels (indicated by a dotted line again).

\begin{figure}\hspace{5em}
\includegraphics[height=17em]{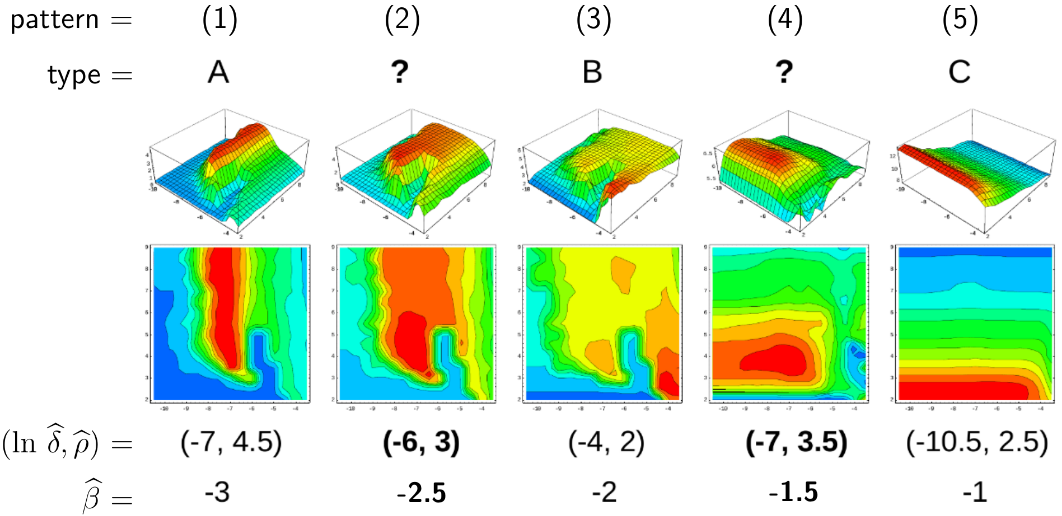}
\caption{\label{fig_patterntypes}
Static pattern versus infection types.
For five transmission rates $\widehat\beta$, fitness landscapes $R_0(\delta,\rho)$ are plotted over pathogen space (3D and via contour) and associated with static patterns and infection types. For two of the transmission rates (bold), the infection type is not clearly differentiated (between A--B and B--C, marked by ``?''). They likely correspond to the static patterns (2) and (4).  
}
\end{figure}

To associate the five static patterns and the three infection types in a more conceivable way,
we have re-computed the fitness landscapes over pathogen space (Fig.~\ref{fig_LF}, top row) for two more transmission rates (Fig.~\ref{fig_patterntypes}).
Those then correspond to the two remaining static patters, even if it turns out to be difficult to associate these extra landscapes with exactly one of our three infection types.
Nevertheless, the transmission rate $\widehat\beta$ is seen again to be a natural parameter here.

\subsection{Natural parameter space}

\begin{figure}\hspace{4em}
(a)\hspace{-.4em}
\includegraphics[height=13em]{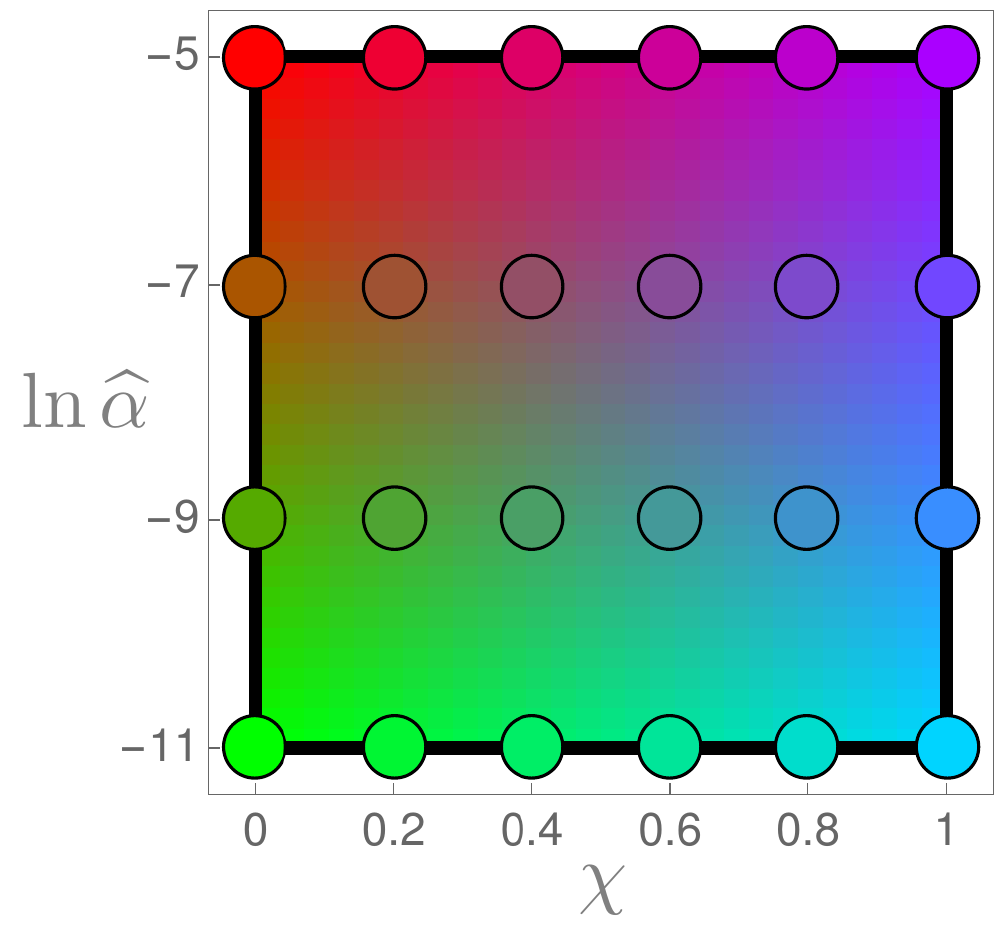}
\hspace{.5em}
(b)\hspace{-.4em}
\includegraphics[height=16em]{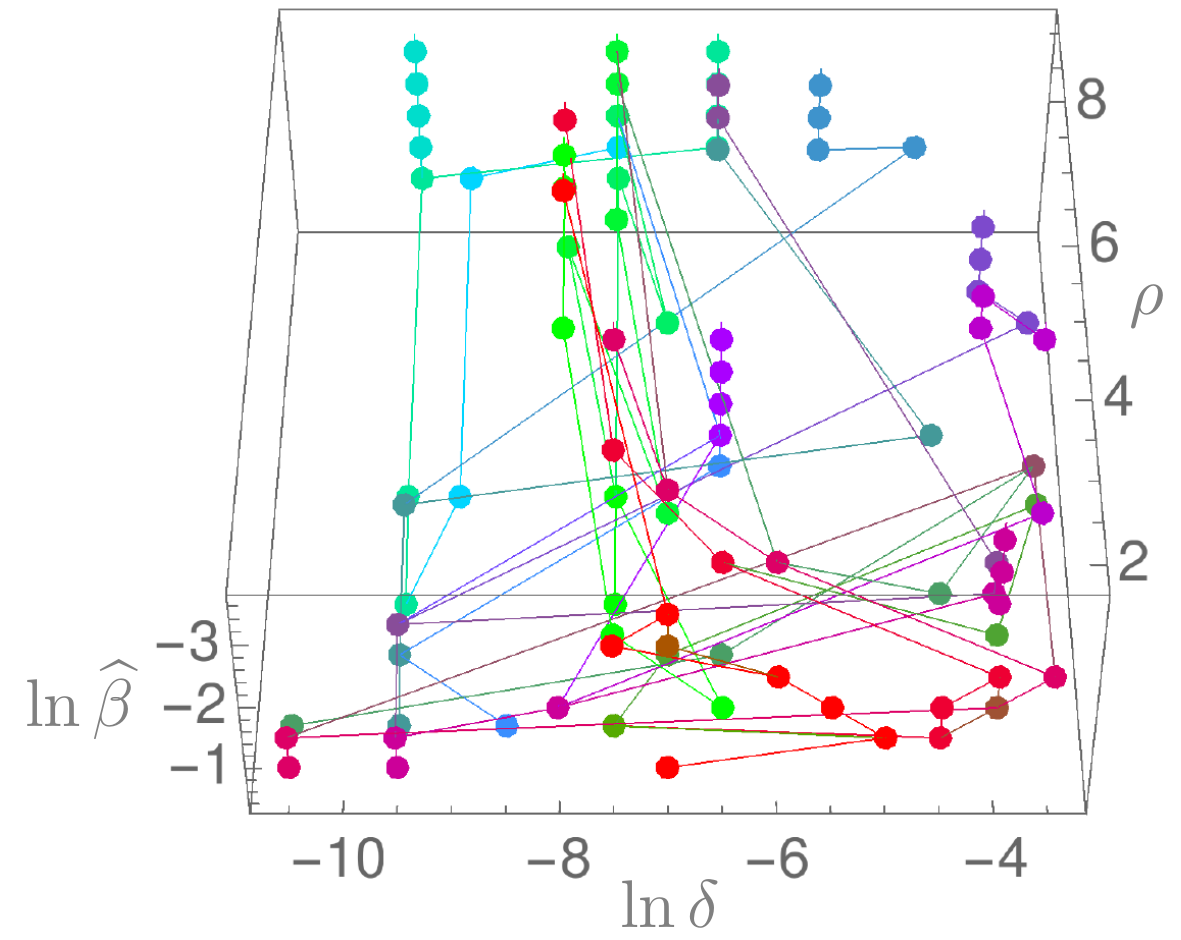}
\caption{\label{fig_3d}
Parameter 
space. Panel (a) defines hue values ($\text{red}\sim\widehat\alpha$, $\text{green}\sim1/\widehat\alpha$, $\text{blue}\sim\chi$) that uniquely depend on cross-immunity $\chi$ and the infectiousness bound $\widehat\alpha$. Panel (b) illustrates the mapping \eqref{map_chidelta} from parameter- to pathogen space; points of the same color---representing the same parameter values $(\chi,\widehat\alpha)$---are connected by a thin line. 
}
\end{figure}

In addition to the transmission rate $\widehat\beta$, it is beneficial to also examine the dependence of the viral fitness on cross-immunity $\chi$ and on the infectiousness bound $\widehat\alpha$.
Here we study the mapping
\begin{subequations}
\label{map_chidelta}
\begin{align}
(\chi,\widehat\alpha)\mapsto
\left(\widehat\delta(\chi,\widehat\alpha;\widehat\beta),
\widehat\rho(\chi,\widehat\alpha;\widehat\beta);\widehat\beta\right)\,,
\end{align}
illustrated in Figure \ref{fig_3d}b, which assigns 
values of the two parameters $(\chi,\widehat\alpha)$---encoded by color (Fig.~\ref{fig_3d}a)---to points in pathogen space that maximize $R_0$,  
\begin{align}
\big(\widehat\delta,\widehat\rho\big)(\chi,\widehat\alpha;\widehat\beta)&=\underset{(\delta,\rho)}{\text{arg\,max}}~\widehat R_0(\chi,\widehat\alpha;\widehat\beta)\,,
\end{align}
\end{subequations}
where $\widehat R_0(\chi,\widehat\alpha;\widehat\beta)=\max_{\alpha\le\widehat\alpha}\,\max_{\delta,\rho}\,R_0(\chi,\widehat\alpha;\widehat\beta,\alpha;\delta,\rho)$.
The dependence on the transmission rate $\widehat\beta$ is captured by a third dimension, erected over pathogen space $(\delta,\rho)$. 

\begin{figure}\hspace{.5em}
(a)\hspace{-.4em}
\includegraphics[height=14em]{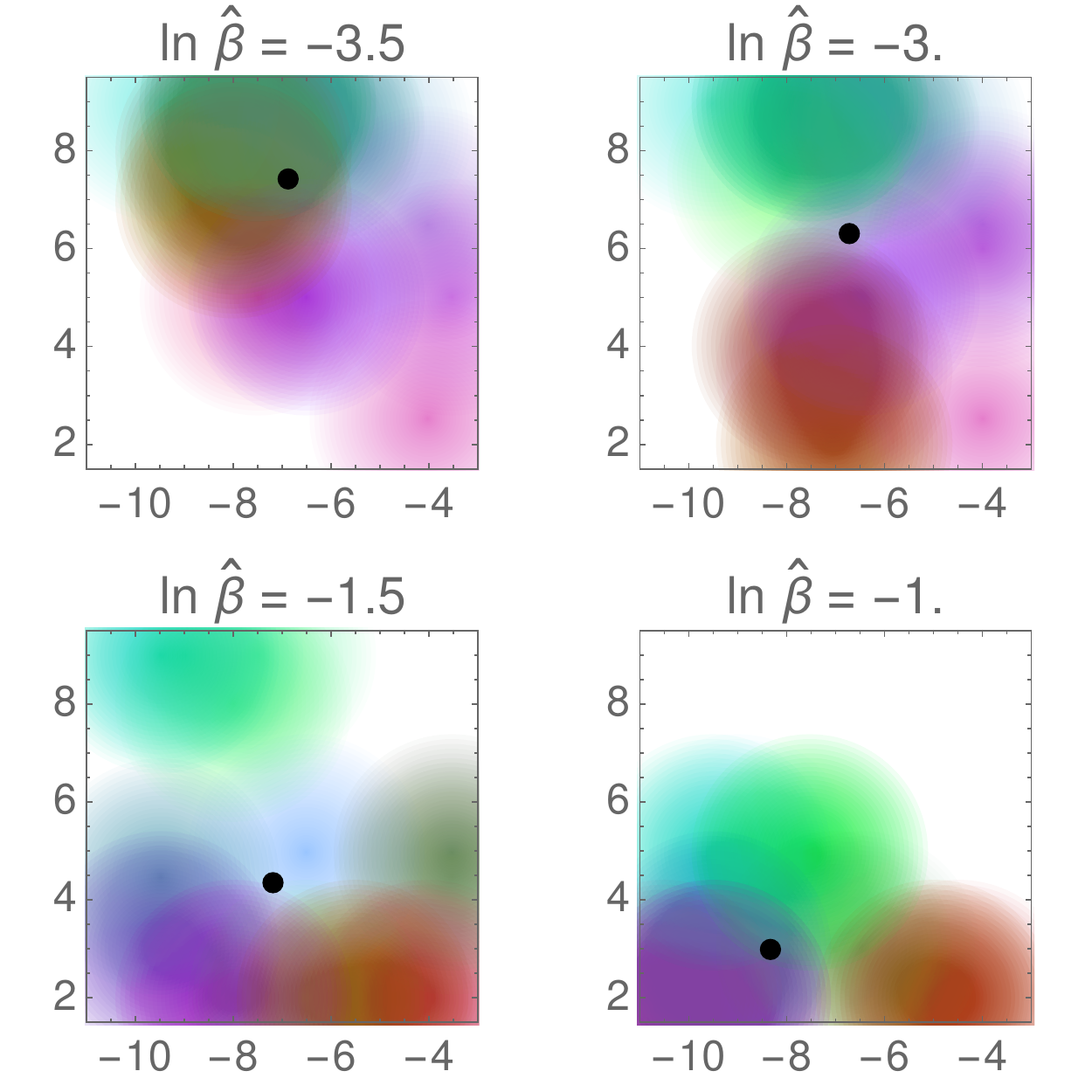}
\includegraphics[height=14em]{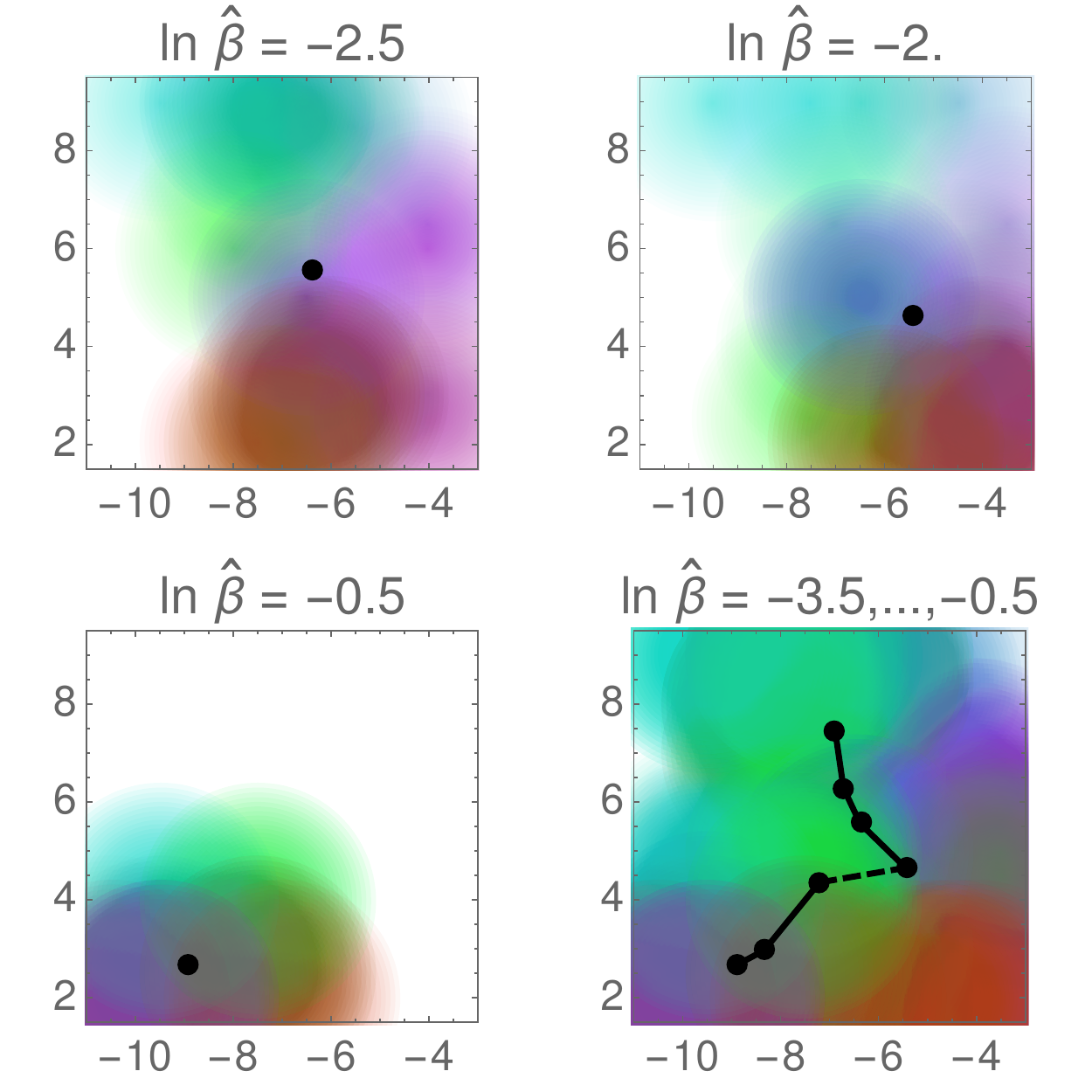}
(b)\hspace{-.4em}
\includegraphics[height=13.2em]{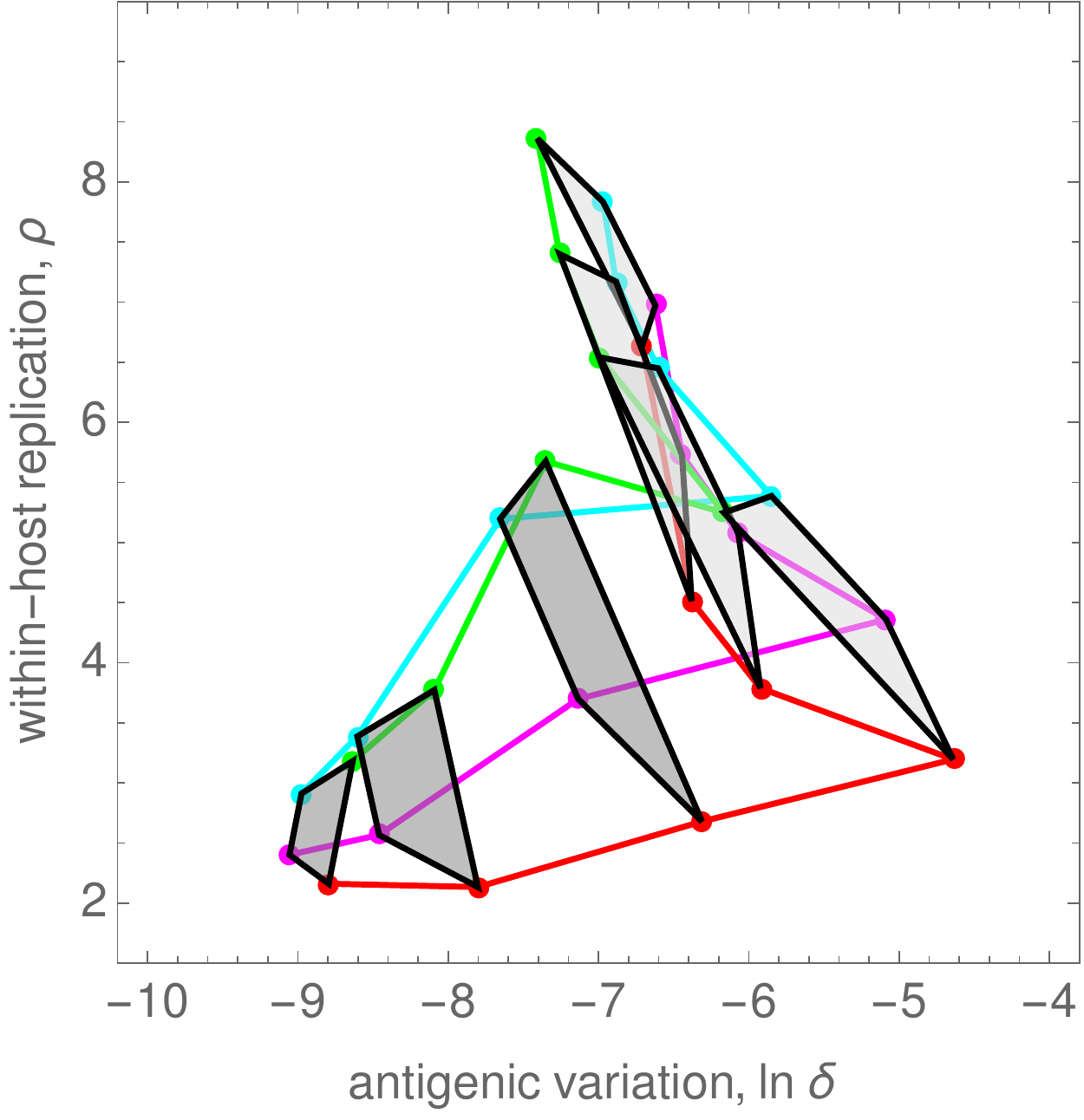}
\caption{\label{fig_extrapolation}
Extrapolation and extremal parameter pairs.
The eight panels in (a) show the fitness maxima in pathogen space for seven transmission rates $\widehat\beta$ and a cumulative combination of them; the colors uniquely represent parameter pairs $(\chi,\widehat\alpha)$ as defined in Figure \ref{fig_3d}a.
The average $(\widehat\delta,\widehat\rho)$-values---taken over the $6\times4$ parameter pairs $(\chi,\widehat\alpha)$---are indicated by black dots; in the cumulative panel, they are connected by black lines.
Panel (b) shows intensity-weighted average locations of the four extreme parameters pairs $(\chi,\widehat\alpha)$ in pathogen space (red, green, violet, cyan in Fig.~\ref{fig_3d}a).
Each of the seven quadrilaterals corresponds to one contact rate ($\ln\,\widehat\beta=-3.5,\dots,-0.5$). The quadrilaterals change their orientation about halfway, when the fitness maxima over $\widehat\beta$ show a discontinuity (cf.~Fig.~\ref{fig_static_patterns}).
}
\end{figure}

Numerical simulations for our (relatively large) parameter space, which cover the within-host dynamics and the transmission network, are hugely time-consuming. They restrict the parameter pairs $(\chi,\widehat\alpha)$---feasible to consider---to be a small number ($=6\times4$).
\footnote{On a usual PC, the simulation runs then take about one night.}
Instead of enlarging this number by increasing the computation power/time,
we decided to proceed by locally extrapolating the simulation results. 
That is, we blur the image points of the mapping \eqref{map_chidelta} by ``enlarging'' these points,
so that they become colored circles. At the same time we decrease the intensity of their unique color towards outer radii.
As a consequence, colors of nearby circles mix according to their red-green-blue content, and we obtain colored patches in pathogen space where the color content corresponds to a unique
$(\chi,\widehat\alpha)$-parameter combination.
The result of that extrapolation is shown in Figure~\ref{fig_extrapolation}a.

Alternatively,
complementing the extrapolation,
we examine the most extreme $(\chi,\widehat\alpha)$-parameter combinations, the corners in Figure \ref{fig_3d}a.
Here one makes an interesting observation; see Figure \ref{fig_extrapolation}b.
The discontinuity between the patterns 3 and 4 (cf. Fig.~\ref{fig_static_patterns})
results in a change of orientation:
\begin{subequations}\label{seq_jump}
\begin{align}
\text{for the patterns}~
\begin{cases}1,2,3\\4,5\end{cases},~
\text{which correspond to}~
\begin{cases}\text{low}\\\text{high}\end{cases}
\text{transmission rates $\widehat\beta$},\\
\text{high values of cross-immunity $\chi$ lie at}~
\begin{cases}(\text{high},\text{high})\\(\text{low},\text{low})\end{cases}
\text{values of $(\delta,\rho)$}\,.
\end{align}
\end{subequations}
In contrast, the values of the infectiousness bound $\widehat\alpha$ that maximize viral fitness do not jump in pathogen space:
high values of $\widehat\alpha$ always lie at $(\text{high},\text{low})$ values of $(\delta,\rho)$.

\begin{figure}\hspace{.5em}
(a)\hspace{-.4em}
\includegraphics[height=14em]{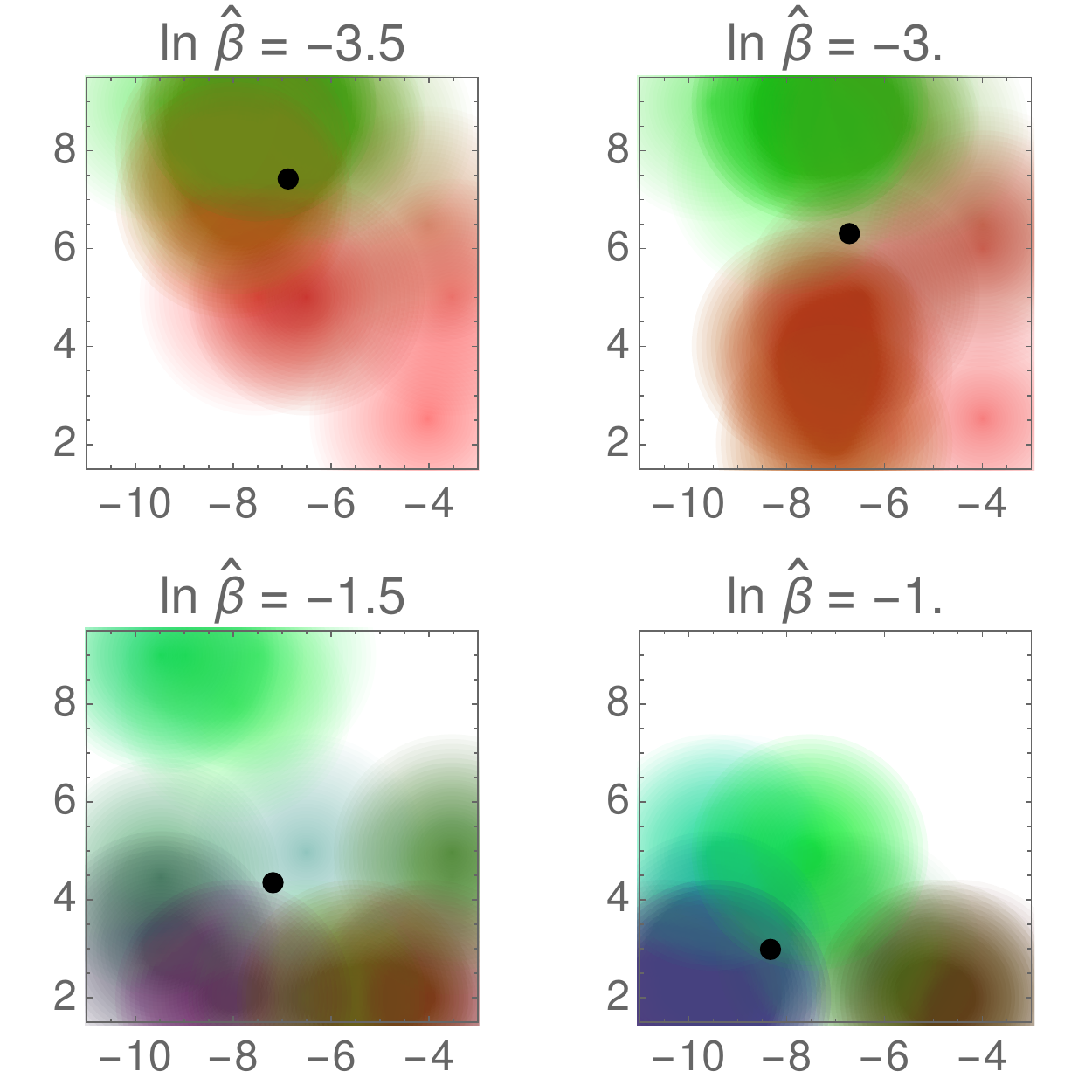}
\includegraphics[height=14em]{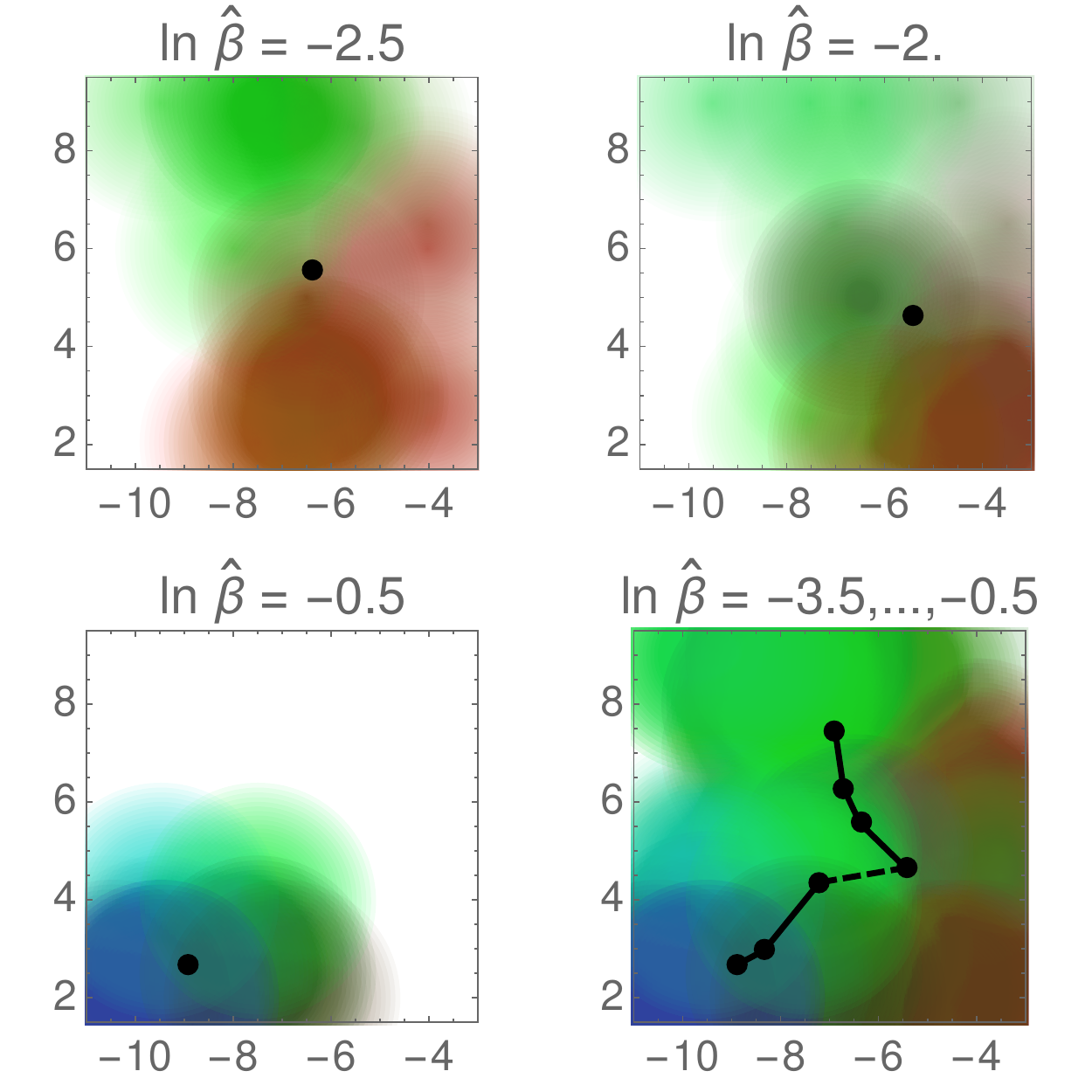}
(b)\hspace{-.4em}
\includegraphics[height=14.3em]{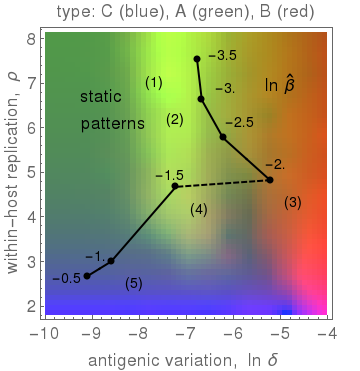}
\caption{\label{fig_typereco}
Infection type reconstruction. The eight panels in (a) show the fitness maxima in pathogen space for seven transmission rates $\widehat\beta$ and a cumulative combination of them; the colors uniquely represent parameter pairs $(\chi,\widehat\alpha)$ as defined by the correspondences \eqref{eq_rgb}.
The average $(\widehat\delta,\widehat\rho)$-values (taken over all colors) are indicated by black dots,
which in the cumulative panel are connected by lines.
Panel (b) shows the infection types A,B,C (colored blue, red, green, resp.) over pathogen space as well as the static patterns (1,\dots,5) and the transmission rate $\widehat\beta$;
its color distribution is approximated well by the cumulative diagram ``$\ln\,\widehat\beta=-3.5,\dots,-0.5$'' in (a).
}
\end{figure}

By red-green-blue mixing of colors (i.e., by forming linear combinations of the parameter content)
the results above can be used to reconstruct the infection types of \cite{LF09}
in terms of three modeling parameters, $\chi,\widehat\alpha,\widehat\beta$.
By their values and particular combination, these parameters represent {\em lifestyles}.
Starting with the color definitions based on antigenic variation (cf.~Fig.~\ref{fig_max_fit}), 
we propose the following simple dependencies,
\footnote{More refined relations can certainly be found.}
\begin{subequations}\label{eq_rgb}
\begin{align}
\text{fitness of type B}&\,\propto\,\widehat\alpha\,,\\
\text{fitness of type A}&\,\propto\,1/(\widehat\alpha\cdot\widehat\beta)\,,\\
\text{fitness of type C}&\,\propto\,\widehat\beta\cdot\chi\,,
\end{align}
\end{subequations}
where $\chi,\widehat\alpha$ contribute hue values as seen in Figure~\ref{fig_extrapolation}a and defined in Figure \ref{fig_3d}a, and
where $\widehat\beta$ provides an intensity weight in accordance with \eqref{seq_jump}.
The resulting color distribution, i.e., the ``mixture'' of lifestyles over pathogen space, is shown in Figure~\ref{fig_typereco};
the similarity of the color content in (a) and (b)---corresponding to
the right- and left hand side expressions in \eqref{eq_rgb}, respectively---is clearly visible.

How does it work? 
The infectiousness bound $\widehat\alpha$ (occurring only in Eqs.~\ref{eq_rgb}a and b)
selects between the types A and B:
if low (i.e., if high loads are required for transmission), type A (i.e., FLI) is favored;
if high (i.e., if low viral loads are sufficient),
type B (i.e., STI) is favored.
According to \eqref{seq_jump},
both these types are favored by rather low transmission rates $\widehat\beta$.
Cross-immunity $\chi$ (scaled blue; cf.~Fig.~\ref{fig_3d}a) favors two patches in pathogen space (cf.~Fig.~\ref{fig_extrapolation}a).
The one with high transmission rates $\widehat\beta$ corresponds to type C (i.e., ChD),
the other we do not really know. It might represent vector-born infections\cite{LF09},
but it is not type C.
Fortunately, this does not matter as in Figure \ref{fig_typereco}a the blue color is switched off at low transmission rates $\widehat\beta$ (cf.~Eq.~\ref{eq_rgb}c).
If $\widehat\alpha,\chi$ are kept fixed, as in Section~\ref{sec_r1}, only the transmission rate selects the infection type in \eqref{eq_rgb}: A for low-, C for high-, and hence, B for medium $\widehat\beta$-values.

\section{Discussion}

Summarizing these last results, we have proposed a mathematical framework equipped with various 
sets of parameters that allows for predicting different realistic types and lifestyles of viral pathogens.
Types refer to the antigenic variation, lifestyles to the evolutionary strategy and corresponding parameter values (cf.~Figs.~\ref{fig_LF} and \ref{fig_patterntypes}).
The parameter sets---including the so-called pathogen- and transmission spaces---cover intra- and inter-host dynamics, including a simple host-contact/transmission network.
Three parameters are necessary for the reconstruction of the observed types/lifestyles: the infectiousness bound $\widehat\alpha$ and the transmission rate $\widehat\beta$, which restrict the possible mode of transmission, and the cross-immunity parameter $\chi$.
Eqs.~\eqref{eq_rgb} establish the fitness definition (Fig.~\ref{fig_typereco}a) for the three infection types of \cite{LF09} (cf.~Fig.~\ref{fig_typereco}b).

Furthermore, referring to the results earlier in the paper, 
we have given an epidemiological interpretation of the static patterns in the phylodynamic theory of Grenfell et al.~\cite{GPGWDMH04}.
Here we claim that the transmission rate $\widehat\beta$ is of particular importance.
By only adjusting its value, transitions between the five static patterns and, correspondingly, the three lifestyles are possible.
Explicitly, this means that the transmission rate and, more general, the contact behavior effectively determine the lifestyle of the pathogen.
The transmission rate $\widehat\beta$ offers a natural (epidemiological)
parametrization of the hand-sketched parabola by Grenfell et al. 
The similarity between that parabola (Fig.~\ref{fig_G04}) and the transmission rate curve 
$\widehat\beta(\widehat\delta,\widehat\rho)$ in Figure~\ref{fig_static_patterns}---obtained strictly by the numerical methods outlined in Section~\ref{sec_meth}---is striking.

Despite these promising first results, there are many ways in which our approach can be improved.
Color-mixing, for example, as utilized for the reconstruction of lifestyles,
is sufficient when dealing with three parameters and three infection types.
For larger numbers, as required in more detailed settings (cf.~Fig.~\ref{fig_patterntypes}),
one needs other tools.
Although, even if less intuitive then,
one could keep the finite value approximation and modify the linear algebra behind.

More parameters and dimensions would come into play when considering:
\begin{enumerate}[(i)]
\item a more involved and tunable network model with multiple/intermediate hosts
\cite{RK03, LHR_08, HCK16}, including indirect transmissions via vectors, air, water, foot, or smear infection \cite{FDA99,SHJ12};
\item further parameters (not only $\delta,\rho$) to be varied by mutation, most importantly infectiousness $\alpha$ \cite{HSL_12}; 
\item reassortment \cite{FGM_13}, possibly as a combination of (i) and (ii); 
\item longer durations of infection (via lower load thresholds $v_0$, fading immunity, etc.),
which would allow for more diverse chronic infections \cite{KH05};
\item virulence \cite{AHMB09} and even longer time scales when co-evolution becomes important \cite{L96,R09};
\item a variable initial viral dose/load \cite{LH14} and
the phenomenon of T-cell exhaustion \cite{WBMMA03, WK15}.
\end{enumerate}
Most of the suggested extensions will not be easy to realize within the presented framework. They would, however, even if implemented partially, substantially improve our understanding of viral evolution.

\section*{Acknowledgments}
This work started towards the end of my stay at Imperial College London, which was funded by the Howard Hughes Medical Institute. I am grateful to Neil Ferguson for many inspiring discussions.

\end{document}